\documentclass[aps,prx,notitlepage,twocolumn,superscriptaddress,longbibliography,nofootinbib]{revtex4-1}
\pdfoutput=1
\usepackage[utf8]{inputenc}
\usepackage[normalem]{ulem}
\usepackage[colorlinks = true,
            linkcolor = blue,
            urlcolor  = blue,
            citecolor = blue,
            anchorcolor = blue]{hyperref}
\usepackage[]{units}
\usepackage{dcolumn,float,tabularx,array,subfigure,booktabs,scalerel,bbold,bm,bbm,braket,color,xcolor,makecell,enumitem,upgreek,blindtext,graphics,verbatim,algorithm,slashed,verbatim,multirow}
\usepackage[noend]{algpseudocode}
\usepackage[bb=boondox]{mathalfa}
\usepackage{amsmath,amsfonts,amssymb,amsthm,mathrsfs,dsfont,mathtools,resizegather}
\usepackage[export]{adjustbox}

\begin{document}

\title{{Designs without disorder: unitary $k$-designs from a single-site bulk defect}}

\author{Samudra Sur\,\,\href{https://orcid.org/0000-0002-4573-4038}
{\includegraphics[scale=0.05]{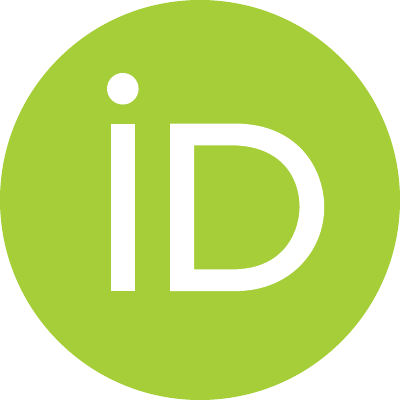}}}
\email{samudra.sur@unige.ch}
\affiliation{Department of Quantum Matter Physics, University of Geneva, 24 quai Ernest-Ansermet, 1211 Genève, Switzerland}
\author{Pratik Nandy\,\,\href{https://orcid.org/0000-0001-5383-2458}
{\includegraphics[scale=0.05]{orcidid.pdf}}}
\email{pratik.nandy@vub.be}
\affiliation{Theoretische Natuurkunde, Vrije Universiteit Brussel (VUB) and \\The International Solvay Institutes, Pleinlaan 2, B-1050 Brussels, Belgium}
\affiliation{RIKEN Centre for Interdisciplinary Theoretical and Mathematical Sciences (iTHEMS),
Wako, Saitama 351-0198, Japan}

\begin{abstract}

We show that an $O(1)$ local defect is sufficient to generate approximate unitary $k$-designs from an otherwise integrable quantum many-body system. Using a two-Pauli-kick (2PK) protocol with temporal sampling, we demonstrate this mechanism in a fixed-magnetization sector of deterministic XXZ spin chains with a single integrability-breaking bulk defect. We derive an analytical selection rule for the Pauli kicks and show that symmetry-preserving strings with a specific domain-wall count, fixed by the couplings, govern design formation. An exhaustive search over all admissible Pauli strings identifies a broad family of kicks that produce designs in the chaotic regime, whereas no such kicks exist in the integrable limit, including for boundary defects. We further show that the design-forming mechanism remains effective for weak bulk defects as the system size increases.

\end{abstract}

~~~~~~~~~~~~~~~~~~~~~~~~~~~~~~~~~~ RIKEN-iTHEMS-Report-26

\maketitle


\section{Introduction}

A unitary $k$-design \cite{Gross:2007xgw, Emerson:2003www} reproduces the first $k$ moments of the Haar measure \cite{Mele:2023ojv} and provides a practical alternative to a Haar-random unitary in randomized benchmarking \cite{Magesan:2010tmc, Eisert:2020ckk}, shadow tomography \cite{Huang:2020tih}, and demonstrations of quantum advantage \cite{Arute:2019zxq}. Most constructions of approximate unitary designs introduce randomness either through local gates \cite{Brandao:2016glh, Hunter-Jones:2019lps, Brandao:2019sgy, Schuster:2024ajb, Hearth:2023hai, Mitsuhashi:2023xdx}, projective measurements \cite{Choi:2021npc, Ho:2021dmh, Cotler:2021pbc}, or Hamiltonian evolution and quenches \cite{Cotler:2017jue,  Chenu:2017qdv, Chenu:2018spm, Zhou:2025noh, Cui:2025teh, Zhou:2026ubz, Sun:2026wrk, Hou:2026cyf, Yang:2026opw, Tian-GangZhou:2026hbi}, with the number of random operations typically growing with system size or circuit depth \cite{Brandao:2016glh, Roberts:2016hpo, Schuster:2024ajb}.

Recently, one of us introduced the two-Pauli Kick (2PK) protocol \cite{Nandy:2026zci}, where neither random Hamiltonians nor a Hamiltonian ensemble is necessary to generate an approximate unitary $k$-design. The protocol uses a fixed Hamiltonian and a fixed Pauli string, with randomness introduced solely through sampling the evolution times from a temporal ensemble. This was demonstrated for single realizations of Gaussian random matrices and Sachdev-Ye-Kitaev Hamiltonians, as well as for the non-integrable limit of the mixed-field Ising chain. Since the underlying Hamiltonians in these examples are either chaotic or break integrability extensively, it remains unclear whether extensive integrability breaking is necessary for design formation. We therefore ask whether an $O(1)$ local perturbation of an otherwise integrable model can generate a unitary $k$-design. If so, this would provide a simple route to generating unitary designs in many-body systems \cite{Vermersch:2018cke, Elben:2018xmb, Joshi:2020quh, Keenan:2022bzo, Joshi:2023rvd}.

To address this question, we consider the integrable spin-$1/2$ XXZ chain, which possesses extensive conserved charges and an exact $U(1)$ symmetry, and is solvable \cite{Orbach:1958zz, Yang:1966ty} by Bethe Ansatz \cite{Bethe_Ansatz}. Rather than an extensive $O(L)$ perturbation \cite{Gubin:2012pxo, Szasz-Schagrin:2021pqg}, we introduce a single-site defect in the bulk. This local perturbation has $O(1)$ operator support, which remains finite in the thermodynamic limit. The defect breaks integrability and produces Wigner--Dyson level statistics \cite{Santos:2004wzv, Barisic, Brenes:2018drq, Lea_Speck_of_Chaos}, with eigenstates obeying the eigenstate thermalization hypothesis (ETH) \cite{Santos_XXZ_ETH, Torres-Herrera:2015ihx, Rigol_XXZ_ETH1, Rigol_XXZ_ETH2}. However, level statistics probe spectral correlations, and ETH \cite{Deutsch:1991quantum, Srednicki:1994chaos, Rigol:2008thermalization, DAlessio:2015qtq} characterizes the matrix elements of local observables in energy eigenstates. A unitary $k$-design probes higher moments of the full distribution of time-evolution operators. Therefore, neither level statistics nor ETH alone is sufficient to establish design formation. Indeed, via the 2PK protocol, we find that the resulting ensemble, generated using the deterministic XXZ Hamiltonian, produces approximate unitary $k$-designs, as quantified by the frame potential \cite{Gross:2007xgw, Roberts:2016hpo}.
The emergence of design depends on the spatial structure of the chosen Pauli kick. For the bulk defect, only a restricted set of symmetry-preserving kicks produces a design in a specific parametric window of ZZ coupling, with their domain-wall structure playing an important role.

In contrast to a bulk defect, the XXZ chain remains integrable in the presence of boundary defects \cite{Cao:2003npb, Sun:2019jpa, Parameshwar:2023prb, Sklyanin:1988yz, Kitanine:2007bi}. By examining the space of allowed kicks, we find that none of these kicks produces an approximate unitary $k$-design. Thus, integrability of the Hamiltonian obstructs design formation, while a single bulk defect is sufficient to remove this obstruction. This remains true for weak bulk defects, over the range studied in this work.

The manuscript is structured as follows. In Sec.\,\ref{sec:model}, we introduce the model. The main protocol for design formation is presented in Sec.\,\ref{sec:2PK}. In Sec.\,\ref{sec:selection_rule}, we study the possible admissible kicks and derive a selection rule for identifying those that can generate designs. Section \ref{sec:results} presents our main results, including an exhaustive analysis of the admissible kicks, finite-size scaling, and the weak-defect limit in the bulk. We conclude with a discussion and outlook in Sec.\,\ref{sec:conclusion}. The supplemental material (SM) contains additional results on design formation and construction for different coupling strengths, a derivation of the selection rule, and a comparison with a related three-step quench protocol (3SP) \cite{Zhou:2026ubz}, which fails to generate designs under the conditions considered here. 


\section{The XXZ chain and defects} \label{sec:model}

We consider the Hamiltonian \cite{Santos:2004wzv,Santos_Mitra_XXZ} with the Pauli operators $\{X_i, Y_i, Z_i\}$:
\begin{align}
H := \sum_{i=1}^{L-1} \left(X_i X_{i+1}
+ Y_i Y_{i+1} + J_{zz} \, Z_i Z_{i+1}\right) + \sum_{j=1}^L h_j Z_j\,,
\label{eq:XXZ}
\end{align}
with open boundary conditions (OBC). Here $J_{zz}$ is the anisotropy parameter and $h_j$ are longitudinal fields (defects). For $h_j=0$, the clean model is Bethe-Ansatz integrable \cite{Bethe_Ansatz}, and it remains integrable if only boundary defects $h_1,h_L\neq0$ are present \cite{Alcaraz:1987uk, Sklyanin:1988yz, Kitanine:2007bi}. However, a single bulk defect $h_j\neq0$ for $j \notin \{1, L\}$ breaks integrability \cite{Santos:2004wzv, Barisic, Brenes:2018drq} (see Supplemental Material (SM) \ref{sm:level_statistics}). We compare two models $H_{\mathrm{bdy}}$ with $(h_1,h_L)=(0.7,1.3)$ and $H_{\mathrm{bulk}}$ with center-defect $h_{\lfloor L/2\rfloor}=1$. Later, we vary the strength and location of the bulk defect to probe their roles in design formation. Throughout the main text, $L=10$ and $J_{zz}=0.9$ (gapless regime). The gapped case ($|J_{zz}|>1$) is treated separately in SM \ref{sm:otherzz}. Irrespective of the position or strength of the defects, the model \eqref{eq:XXZ} conserves total magnetization, \emph{i.e.}, $[H, \sum_i Z_i] = 0$, associated with $U(1)$ symmetry. We work in the zero-magnetization (half-filled) sector of dimension $d=\binom{L}{L/2}$, and assess the design properties in this irreducible block.


\section{Frame potential and 2PK protocol} \label{sec:2PK}

For an ensemble $\mathcal{E}$ of
unitaries acting on a $d$-dimensional Hilbert space, the $k$-fold channel is defined as
$\Phi^{(k)}_{\mathcal{E}}(O)=\mathds{E}_{U\sim\mathcal{E}}
\big[U^{\otimes k}O\,U^{\dagger\otimes k}\big]$, which captures all the $k$-th order moments of the ensemble \cite{Mele:2023ojv}. The ensemble $\mathcal{E}$ is a
unitary $k$-design when its $k$-fold channel matches the $k$-fold channel evaluated over the full Haar ensemble $\Phi^{(k)}_{\mathcal{E}}=\Phi^{(k)}_{\mathrm{Haar}}$. The efficacy of a $k$-design is measured by the $k$-th frame potential of the ensemble \cite{Gross:2007xgw, Roberts:2016hpo}:
\begin{align}
F^{(k)}_{\mathcal{E}}:=\mathds{E}_{U,V\sim\mathcal{E}} \big|\mathrm{Tr} \big(U^\dagger V\big)\big|^{2k}\,,
\label{eq:framepot}
\end{align}
with $F^{(k)}_{\mathcal{E}_{\mathrm{Haar}}}=k!$ for $k \le d$ \cite{Scott_2008}. The normalized quantity $F^{(k)}_{\mathcal{E}}/k!$ equals unity for an exact $k$-design and will be our primary benchmark throughout the manuscript. We refer to an ensemble as an ``approximate $k$-design'' when $F^{(k)}_{\mathcal{E}}/k! \to 1$ in the large $d$ limit.

The 2PK ensemble is defined as \cite{Nandy:2026zci}
\begin{align}
\mathcal{E}_{\mathrm{2PK}}:=\big\{U_{\mathrm{2PK}}=e^{-iHt_3}\,P\,e^{-iHt_2}\,P\,e^{-iHt_1}\big\}\,,
\label{eq:2PK}
\end{align}
where $t_i\sim p(t)$ are independently sampled from a uniform temporal distribution on $[0,T]$. The Hamiltonian $H$ is Hermitian, and the kick $P$ is non-identity, Hermitian, and unitary ($P \neq \mathds{1}, P = P^{\dagger}, P^2 = \mathds{1}$). We keep both $H$ and $P$ fixed and generate the ensemble solely by sampling the evolution times $t_1,t_2$ and $t_3$. For comparison, we also use the one-kick (1PK) ensemble $\mathcal{E}_{\mathrm{1PK}}=\{e^{-iHt_2}Pe^{-iHt_1}\}$, which is the minimal member of the same family of generic $n$-Pauli kick ($n\mathrm{PK}$) protocols \cite{Nandy:2026zci}. For the numerical results, we set the upper bound $T=10^6$ to approximate the perfect-filter
limit $T \to \infty$; this choice is validated numerically below.

For $k=1$, the relevant quantities for the frame potential can be determined by a single matrix $w_{ab}:=\big|\langle E_a|P|E_b\rangle\big|^2$, where $H|E_a\rangle=E_a|E_a\rangle$ within the chosen magnetization sector. By construction, $w$ is real, symmetric, and doubly stochastic, \emph{i.e.}, $\sum_b w_{ab}=1$. The largest eigenvalue of $w$ is unity (Perron eigenvalue), with the corresponding eigenvector being the uniform vector. Assuming a non-resonant (non-degenerate for $k=1$) spectrum of $H$, the $k=1$ frame potential for $n$PK protocol is given by $F^{(1)}_{n\mathrm{PK}}=\mathrm{Tr}\big(w^{2n}\big)$. Defining $\mu_j$ to be the eigenvalues of $w$ in descending order, we have $\mathrm{Tr}(w^{2n})=1+\sum_{j\ge2}\mu_j^{2n} \ge 1$. The equality holds for all $n$ if all $\mu_j =0$ for $j \geq 2$, \emph{i.e.}, if $w_{mn}=1/d$ (the flat matrix) \cite{Nandy:2026zci}. However, in our setting, the $w$ matrix fluctuates. Therefore, we have
\begin{align}
F^{(1)}_{\mathrm{1PK}}=\mathrm{Tr}(w^{2})\equiv\Lambda \,,~  F^{(1)}_{\mathrm{2PK}}=\mathrm{Tr}(w^{4}) \simeq 1+\frac{2(\Lambda-1)^2}{d}\,, \label{eq:F1general}
\end{align}
where $\Lambda$ is the kurtosis factor of the Pauli matrix elements in the eigenbasis of the Hamiltonian, giving $\Lambda = 3$ for real (GOE) and $\Lambda = 2$ for complex (GUE) eigenvectors. For chaotic eigenvector statistics with $\Lambda\ge 2$, the one-kick ensemble can never attain the Haar value $F^{(1)}_{\mathcal{E}_{\mathrm{Haar}}}=1$. The second kick instead allows $\mathrm{Tr}(w^4)\to1$, leading to $F^{(1)}_{\mathrm{ 2PK}} \to 1$, which is required for the design.

For a general $k$ in the large $d$ limit, the 1PK frame potential for a chaotic $H$ with GOE/GUE statistics is given by \cite{Nandy:2026zci}
\begin{align}
F^{(k)}_{\mathrm{ 1PK}}=k!\sum_{j=0}^{k}\binom{k}{j}\,!(k-j)\,\Lambda^{\,j}\,,
\label{eq:F1PKGXE}
\end{align}
where $!m$ denotes the derangement number. Even in the ideal flat-matrix limit $\Lambda=1$ (as $w_{mn} = 1/d$), one finds
$F^{(k)}_{\mathrm{1PK}}=(k!)^2$, exceeding the Haar value $k!$ for
$k\ge2$. Hence, the one-kick ensemble cannot form a $k$-design even for
an ideally delocalized kick.

For 2PK, however, the intermediate evolution $e^{-iHt_2}$ in \eqref{eq:2PK} randomizes the effective coupling matrix itself. The leading deviation from Haar at $k=1$ is \cite{Nandy:2026zci}
\begin{align}
\Delta^{(1)}_{\mathrm{2PK}} := F^{(1)}_{\mathrm{2PK}} - 1 = \mathrm{Tr}(w^{4}) -1 \simeq \frac{2(\Lambda-1)^2}{d}\,, \label{eq:Delta1RMT}
\end{align}
\emph{i.e.}, $d\,\Delta^{(1)}_{\mathrm{2PK}}\to8$ for GOE ($\Lambda=3$). Under the factorization $F^{(k)}_{\mathrm{2PK}}=k!\,[\mathrm{Tr}(w^4)]^k[1+O(1/d)]$ established in Ref.\,\cite{Nandy:2026zci},
the entire $k$-dependence is fixed by $k=1$ value of $\mathrm{Tr}(w^4)$. Therefore,
$\Delta^{(1)}_{\mathrm{2PK}}\to0$ implies $F^{(k)}\to k!$ for all $k$ for $k^2 \ll d$. The factorization itself rests on the non-resonance assumption, which we verified to hold by comparing with the sampled frame potentials up to $k=5$ in all examples we studied. Because $\Delta^{(1)}_{\mathrm{2PK}}$ and $\Lambda \equiv \mathrm{Tr}(w^2)$ are obtained directly from a single diagonalization without temporal sampling, we use them as exact diagnostics in the exhaustive kick search and in our finite-size and disorder-dependence analyses.

\begin{figure}[t]
    \hspace{-0.5cm}
\includegraphics[width=0.77\linewidth]{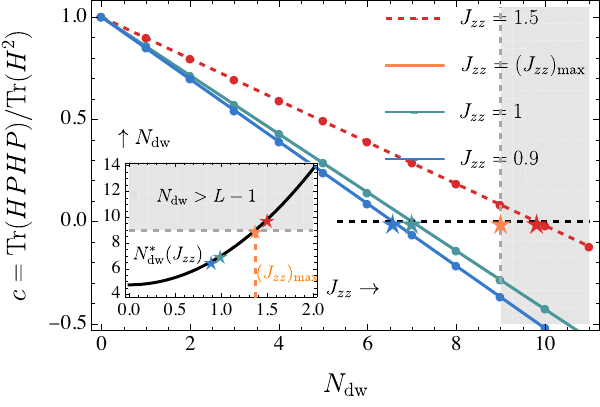}
    \caption{Selection rule for the bulk-defect model $(L = 10, h_{\lfloor L/2 \rfloor} = 1)$. \textbf{Main:} overlap ratio $c$ of Eq.\,\eqref{c_Delta_Ndw} versus domain-wall number $N_{\mathrm{dw}}$; circles mark realizable integer $N_{\mathrm{dw}}$, stars denote the roots of $c = 0$ (\emph{i.e.}, $N^{*}_{\mathrm{dw}}$) from \eqref{Ndwstar}, and the gray shaded area to the right denotes the inaccessible region $N_{\mathrm{dw}} > L-1$. At $J_{zz} = 1$, the root falls exactly on $N_{\mathrm{dw}}^{*} = 7$ (green star); at $J_{zz} = 1.5$ (red star) it lies at $9.81$. \textbf{Inset:} the optimal count $N^{*}_{\mathrm{dw}} (J_{zz})$ (solid black line), which reaches $L-1$ at $(J_{zz})_{\max} = 1.374$, Eq.\,\eqref{Deltamax}.}
    \label{fig:kickanalyticplot}
\end{figure}


\section{Selection rule and admissible kicks} \label{sec:selection_rule}

We first need to determine the space of admissible kicks. Since we work on fixed-magnetization sector, only Pauli strings $Z$ and the identity are considered: $P_S = \prod_{i \in S} Z_i$ with $S \subseteq \{1, \cdots, L\}$. As an example, for $L = 6$ and $S = \{1, 3, 5\}$, we have $P_S = Z_1 Z_3 Z_5$ $\equiv$ ZIZIZI. Pauli strings containing $X$ or $Y$ are excluded, even though some of their combinations can preserve the total magnetization for some basis states. However, they do not, in general, preserve the fixed-magnetization sector for all states. Restricting to $Z$ and $I$ only, there are $2^L$ possible subsets $S$, but $S$ and its complement $\bar{S}$ generate equivalent kicks within a fixed magnetization sector: $P_{\bar{S}}
=\left(\prod_{i=1}^L Z_i\right)P_S$ and $\prod_i Z_i$ acts as a fixed phase on every state in the sector. Hence, the total $2^L$ subsets reduce to $2^{L-1}$ inequivalent strings. Excluding the class containing the trivial identity leaves $2^{L-1}-1$ nontrivial admissible kicks.

To characterize their spatial structure, let $N_{\mathrm{dw}}$ denote the number of domain walls in the string. A domain wall occurs whenever two neighboring sites have different entries, $\{Z,I\}$ and $\{I,Z\}$. For example, the string  ZIIIZIZIZZ has $N_{\mathrm{dw}} = 6$. Therefore, we can write
\begin{align}
N_{\mathrm{dw}}
= \sum_{i=1}^{L-1}
\mathbf{1} \left[
\left|S\cap\{i,i+1\}\right|=1
\right]\,, \label{Ndw_def}
\end{align}
where $\mathbf{1}$ is the indicator function and $| \cdot |$ is the cardinality of the set. Since there are $L-1$ neighboring bonds, $(N_{\mathrm{dw}})_{\max}=L-1$. This maximum is reached only when every neighboring pair alternates, giving the two N\'eel patterns
ZIZI... and IZIZ..., which belong to the same equivalence class. Under conjugation by a $Z$-string, the three families of terms in Eq.\,\eqref{eq:XXZ} behave differently:
\begin{align*}
&P X_i X_{i+1} P = (-1)^{\left|S\cap\{i,i+1\}\right|} X_i X_{i+1}, ~P Z_j P = Z_j\,, \nonumber \\
&P Y_i Y_{i+1} P = (-1)^{\left|S\cap\{i,i+1\}\right|} Y_i Y_{i+1}, ~ P Z_i Z_{i+1} P = Z_i Z_{i+1}\,.
\end{align*}
The sign on an $XX$ or $YY$ bond is $-1$ precisely when that bond is a
domain wall of the string. Since distinct Pauli strings follow the trace-orthogonality relation, no cross terms survive in the overlap between $H$ and $PHP$, \emph{i.e.}, $\operatorname{Tr}(H P H P)$. The overlap ratio $c \equiv c(J_{zz}, N_{\mathrm{dw}})$ introduced in \cite{Nandy:2026zci} quantifies the first-order overlap between $H$ and $PHP$ and provides a simple criterion for excluding a set of kicks. For the XXZ chain, it can be computed analytically as (see SM \ref{sm:derv})
\begin{align}
c := \frac{\operatorname{Tr}(H P H P)}
{\operatorname{Tr}(H^2)}
= 1-  \frac{4N_{\mathrm{dw}}
}{(2+J_{zz}^2)(L-1)+\sum_j h_j^2
}\,,\label{c_Delta_Ndw}
\end{align}
providing a selection rule for the admissible kicks in the XXZ chain. The $(L-1)$ factor in the denominator arises from the number of nearest-neighbor bonds in the chain in OBC. Increasing $N_{\mathrm{dw}}$ decreases $c$, thereby increasing the Hilbert–Schmidt distance between $PHP$ and $H$ as
\begin{align}
    d^2_{\mathrm{HS}}(H, PHP) &= \|H\|_{\mathrm{HS}}^2 +  \|PHP\|_{\mathrm{HS}}^2 - 2 \, \mathrm{Tr}(H\,PHP) \nonumber \\
    &= 2 \, \mathrm{Tr}(H^2) (1-c)\,,
\end{align}
where we have used the unitary invariance of the Hilbert –Schmidt norm: $\|H\|_{\mathrm{HS}}^2 =  \|PHP\|_{\mathrm{HS}}^2 = \mathrm{Tr}(H^2)$. Setting $c = 0$ in \eqref{c_Delta_Ndw} gives the optimal domain-wall count (operator-space orthogonality) for a set of parameters:
\begin{align}
    N^{*}_{\mathrm{dw}} (L, J_{zz}, h_j) = \frac{1}{4}\Big[(2+J_{zz}^2)(L-1)+\sum_j h_j^2\Big]\,. \label{Ndwstar}
\end{align}
Therefore, the selection rule is determined by the specific spatial structure of the kick, with the required domain-wall count set by $J_{zz}$ and $h_j$ for any fixed system size $L$. Since, even for the optimal choice, $N_{\mathrm{dw}}^{*} \leq L-1$, a solution exists only if
\begin{align}
    J_{zz} \leq (J_{zz})_{\mathrm{max}}
= \sqrt{2-\frac{\sum_j h_j^2}{L-1}}
\xrightarrow[L\to\infty]{} \sqrt{2}\,. \label{Deltamax}
\end{align}
The limit holds provided
$\sum_j h_j^2=O(L^\alpha)$ with $\alpha<1$, and exactly in the field-free limit (see SM \ref{sm:otherzz} for the analysis in the gapped regime and beyond $J_{zz}> (J_{zz})_{\mathrm{max}}$).

\begin{figure}[t]
    \hspace{-0.3cm}
\includegraphics[width=1\linewidth]{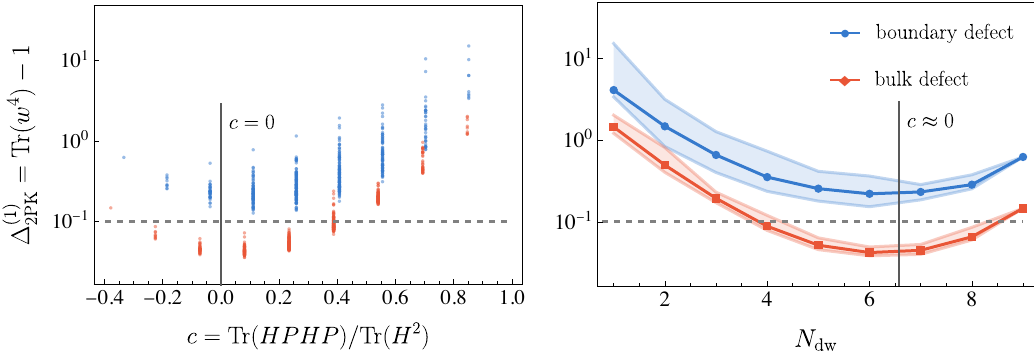}
    \caption{\textbf{Left:} Selection rule for the Pauli kicks obtained from an exhaustive search over all 511 admissible kicks for $L=10$, $J_{zz} = 0.9$, $(h_{1}, h_L) = (0.7,1.3)$ (boundary defect) and $h_{\lfloor L/2 \rfloor} = 1$ (bulk defect). Each point is characterized by the corresponding values of $\Delta_{\mathrm{2PK}}^{(1)}$ and $c$. The dashed line is $10\%$ tolerance above the minimum of $\Delta_{\mathrm{2PK}}^{(1)}$. \textbf{Right:} Variation of $\Delta_{\mathrm{2PK}}^{(1)}$ with the domain-wall number $N_{\mathrm{dw}}$. The solid line shows the median over admissible kicks at fixed $N_{\mathrm{dw}}$, while the shaded band denotes the corresponding 10th--90th percentile range. The vertical dashed line marks $N_{\mathrm{dw}}^{*}$, the domain-wall number for which $c \approx 0$.}
    \label{fig:FP2PKGUEwithbetaplot}
\end{figure}

For $L=10$, $(J_{zz})_{\max}\simeq1.326$ for the boundary-defect
($\sum_j h_j^2=2.18$) and $(J_{zz})_{\max}\simeq1.374$ for the bulk-defect model ($\sum_j h_j^2=1$). For our chosen $J_{zz}=0.9$, the optimal values are $N^{*}_{\mathrm{dw}}\simeq6.9$ and $6.6$ for the boundary and bulk-defect models, respectively. We therefore choose a kick with
$N_{\mathrm{dw}}=6$, close to the predicted optimal kick value $N_{\mathrm{dw}}^{*}$. Table \ref{tab:kicks} and Fig.\,\ref{fig:KICKLISTgap} in the SM summarize the performance of different kicks for both models. Figure \ref{fig:kickanalyticplot} shows the optimal kick selection, together with the corresponding $N_{\mathrm{dw}}$, for several values of $J_{zz}$.


\section{Main Results} \label{sec:results}

\subsection{Exhaustive search and representative kicks} 

To verify Eq.\,\eqref{c_Delta_Ndw} directly, we perform an exhaustive search over all $2^{L-1}-1=511$ admissible kicks at $L=10$. For each kick, we evaluate the first-order deviation
$\Delta_{\mathrm{2PK}}^{(1)}=\mathrm{Tr}(w^4)-1$. Figure \ref{fig:FP2PKGUEwithbetaplot} (left and right panels) shows the result as a function of
$c$ and $N_{\mathrm{dw}}$: the kicks minimizing $\Delta_{\mathrm{2PK}}^{(1)}$
cluster around $c\simeq0$ and, equivalently, around the domain-wall number $N_{\mathrm{dw}}^{*}$ predicted by Eq.\,\eqref{Ndwstar}, confirming the selection rule. The optimum is attained by many kicks (shaded band), so no fine-tuning of a specific string is required.

\begin{figure}[t]
\hspace*{-0.2cm}%
\includegraphics[width=1\linewidth]{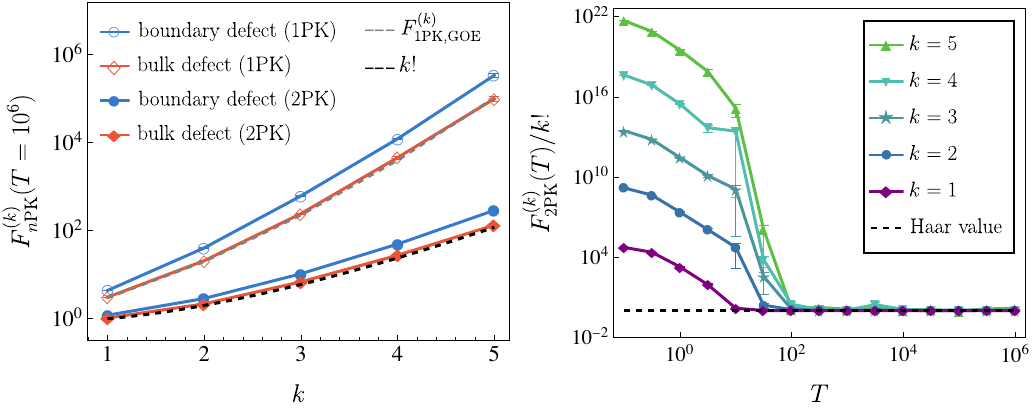}
\caption{\textbf{Left:} 1PK and 2PK Frame potential with boundary and bulk defects, with the Pauli kick $P_1=$ ZIIIZIZIZZ. The gray and black dashed lines denote \eqref{eq:F1PKGXE} with $\Lambda = 3$ (GOE) and the Haar value $k!$, respectively. \textbf{Right:} The normalized $F^{(k)}_{\mathrm{2PK}}/k!$, saturating to unity (black dashed line) for $T \geq 10^4$. Here $L = 10$ with a half-filling sector. The parameters are $J_{zz} = 0.9$ with $(h_1, h_L) = (0.7,1.3)$ for the boundary and $h_{\lfloor L/2 \rfloor} = 1$ for the bulk defects. The results are averaged over $10^5$ independent temporal realizations.}
\label{fig:FPnPKXXZL10Jzz0p9Tplot}
\end{figure}

The left panel of Fig.\,\ref{fig:FPnPKXXZL10Jzz0p9Tplot} shows $F^{(k)}$ for the boundary- and bulk-defect models under 1PK and 2PK at $J_{zz} = 0.9$. For the bulk defect, the 1PK data track the GOE prediction of Eq.\,\eqref{eq:F1PKGXE} across the full range of $k$, while 2PK falls close to the Haar value $k!$. For the boundary defect, both protocols sit above the Haar value and therefore do not form the approximate $k$-design. In all cases, we take $T = 10^6$, which is justified by the observation that the normalized frame potential $F^{(k)}_{\mathrm{2PK}}/k!$ stabilizes close to unity after $T \geq 10^4$ (right panel).

\begin{table}[t]
\hspace{-0.2cm}
\small
\setlength{\tabcolsep}{0.05pt}
\begin{tabular}{cc|ccc|ccc}
\hline\hline
&&\multicolumn{3}{c|}{$H_{\mathrm{bdy}}$}
&\multicolumn{3}{c}{$H_{\mathrm{bulk}}$}\\
$L$ & Kick $P$
& $\mathrm{Tr}(w^2)$ & $\mathrm{Tr}(w^4)$ & $d \Delta_{\mathrm{2PK}}^{(1)}$
& $\mathrm{Tr}(w^2)$ & $\mathrm{Tr}(w^4)$ & $d \Delta_{\mathrm{2PK}}^{(1)}$\\
\hline
8  & ZIIZIZII        & 3.70  & 1.25 &   17.26 & 2.87 & 1.10 &  7.19 \\
10 & ZIIIZIZIZZ      & 4.38  & 1.20 &  49.98 & 3.07 & 1.04 & 8.88 \\
12 & IZIIZIZIZZII    & 5.17  & 1.18 &  166.73 & 3.07 & 1.01 & 8.87 \\
14 & ZIZIIZIZZIZZII    & 5.43 & 1.17 & 572.22 & 3.03 & 1.00 & 8.44 \\
16 & ZZIZIZZIZIIZIZZI    & 5.57 & 1.11 & 1424.47 & 3.02 & 1.00 & 8.42\\
\hline\hline
\end{tabular}
\caption{Exact first-order diagnostics for the boundary ($H_{\mathrm{bdy}}$) and bulk-defect ($H_{\mathrm{bulk}}$) models in the half-filled sector. All of the chosen kicks are near-optimal for the bulk defect and give $c \simeq 0$. The eigenvector kurtosis $\mathrm{Tr}(w^2)$, the exact first-order frame potential $\mathrm{Tr}(w^4)$, and the scaled excess $d\Delta_{\mathrm{2PK}}^{(1)} =d[\mathrm{Tr}(w^4)-1]$, should approach $\Lambda = 3$ (GOE), 1, and $2 (\Lambda - 1)^2 = 8$ as $d$ increases.}
\label{tab:scaling}
\end{table}


\subsection{Scaling with Hilbert-space dimension}

\begin{figure}[t]
    \hspace{-0.5cm}
\includegraphics[width=1.02\linewidth]{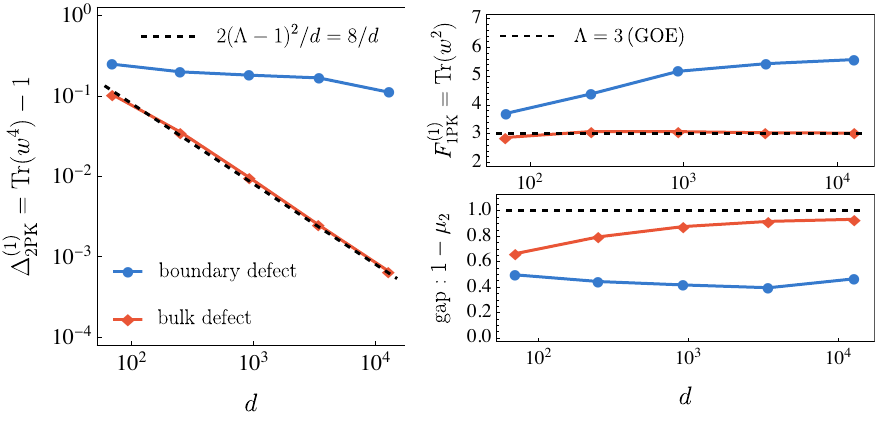}
    \caption{\textbf{Left:} Deviation of the exact first-order 2PK frame potential from Haar value; the bulk defect follows $\Delta_{\mathrm{2PK}}^{(1)} \sim 1/d$ (dashed line: GOE prediction $8/d$), the boundary defect shows slower decay than $1/d$. \textbf{Right, top:} $\mathrm{Tr}(w^2)$ converges to GOE kurtosis $\Lambda = 3$ for the bulk defect and increases for the boundary defect. \textbf{Right, bottom:} The gap $1-\mu_2$ stays open for the bulk defect and stays far from unity for the boundary defect as $d$ increases.}
    \label{fig:Deltaw2vsdplotXXZ}
\end{figure}

Table \ref{tab:scaling} and Fig.\,\ref{fig:Deltaw2vsdplotXXZ} show the finite-size scaling, namely whether design formation survives as $d$ increases ($L = 8,10,12,14,16$). For the bulk-defect model, the first-order deviation follows $\Delta_{\mathrm{2PK}}^{(1)} \sim 1/d$, consistent with the GOE prediction $8/d$ [Eq.\,\eqref{eq:Delta1RMT}], and the eigenvector kurtosis $\mathrm{Tr}(w^2)$ converges to $\Lambda=3$. The scaled excess $d\,\Delta^{(1)}_{\mathrm{2PK}}$ remains close to the value $2(\Lambda-1)^2=8$ with $O(1/d)$ corrections. In contrast, for the boundary model, $\Delta_{\mathrm{2PK}}^{(1)}$ exhibits much slower decay than $1/d$, while $\mathrm{Tr}(w^2)$ increases over the accessible sizes, without saturating at the GOE value. These quantities follow from a single diagonalization, without any temporal sampling.

We can use a single diagnostic to capture both $\Delta^{(1)}_{\mathrm{2PK}}$ and $\mathrm{Tr}(w^2)$. Since $w$ is doubly stochastic, its largest eigenvalue is unity (Perron eigenvalue), and $\mu_2^4\le\Delta^{(1)}_{\mathrm{2PK}}\le\mu_2^2[\mathrm{Tr}(w^2)-1]$ with $\mu_2 := \mathrm{max}_{j \geq 2} |\mu_j|$ the largest subleading eigenvalue in magnitude  \cite{Nandy:2026zci}. So the emergence of design is equivalent to the opening of the spectral gap $1-\mu_2$ whenever $\mathrm{Tr}(w^2)$ stays bounded. Figure \ref{fig:Deltaw2vsdplotXXZ} (right, bottom panel) shows the two models on opposite sides of this criterion: the gap opens (\emph{i.e.}, approaches unity) with $d$ for the bulk defect while it stays far from unity for the boundary defect. In the integrable chain, a slow mode of $w$ persists at all sizes studied here. Since no admissible kick can remove it, the protocol does not lead to a design.

\subsection{Weak defect and approximate unitary $k$-design}

\begin{figure}[t]
    \hspace{-0.4cm}
\includegraphics[width=1\linewidth]{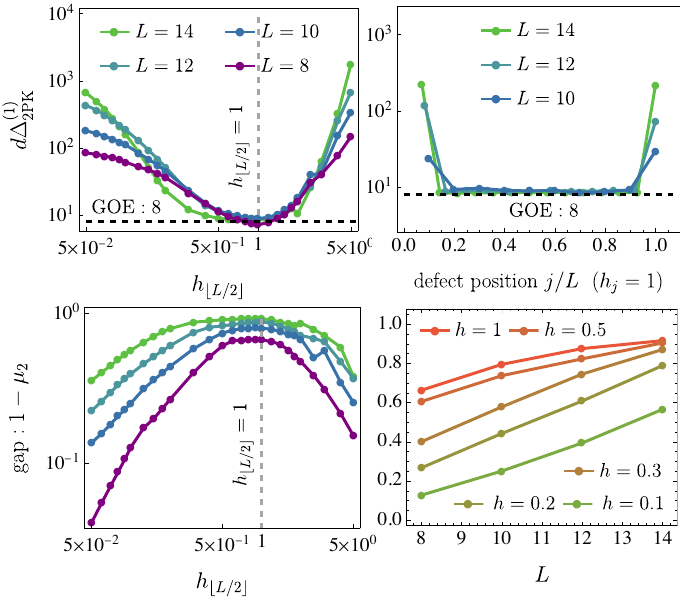}
    \caption{\textbf{Left:} Variation of $d \Delta^{(1)}_{\mathrm{2PK}}$ (dashed line: GOE prediction $2(\Lambda - 1)^2 = 8$) \textbf{(top)} and the gap $1-\mu_2$ \textbf{(bottom)} with the bulk defect strength $h_{\lfloor L/2 \rfloor}$. The gap remains open for higher $L$ with longer windows of the defect strength. \textbf{Right, top:} The deviation $d \Delta^{(1)}_{\mathrm{2PK}}$ with the position of the defect, which follows the GOE prediction at bulk sites but sharply increases at the boundary. \textbf{Right, bottom:} Finite-size scaling of the gap for different defect strength $h \equiv h_{\lfloor L/2 \rfloor}$. Kicks for $L \leq 14$ are taken from Table \ref{tab:scaling}.}
    \label{fig:bulkpositionplot}
\end{figure}

Finally, we examine how weak a bulk defect can be while still generating a unitary $k$-design. Figure \ref{fig:bulkpositionplot} (left) shows two different diagnostics as the defect strength $h \equiv h_{\lfloor L/2\rfloor}$ is varied at fixed optimal kick. Both $d\Delta^{(1)}_{\mathrm{2PK}}$ (top panel) and the gap $1-\mu_2$ (bottom panel) are non-monotonic, indicating that design formation occurs only within a finite window of bulk defect strengths. At fixed $h$, the gap opens monotonically with $d$ for every value we examined, until $h=0.1$ [Fig.\,\ref{fig:bulkpositionplot} (right, bottom)], suggesting that even weak defects may support design formation in the thermodynamic limit. This behavior is consistent with the level statistics analysis in Ref.\,\cite{Lea_Speck_of_Chaos}.
Figure \ref{fig:bulkpositionplot} (right, top) distinguishes the role of the location of the defect: at fixed $h_j=1$, the deviation stays at the GOE value for every bulk site $j$ and sharply increases only when the defect sits exactly at $j=1$ or $j=L$ (boundaries), where integrability is restored, and design formation breaks down. A defect one site inward behaves as a bulk defect, consistent with the level statistics.


\section{Conclusion and outlook} \label{sec:conclusion}

We have shown that a single $O(1)$ local defect and a fixed Pauli kick are sufficient to generate approximate unitary $k$-designs from an integrable many-body Hamiltonian, with randomness coming solely from temporal sampling. The condition $c\simeq0$ corresponds to near-orthogonality of $H$ and $PHP$ in operator space (in terms of Hilbert-Schmidt distance). For longitudinal Pauli strings, this condition reduces to a domain-wall criterion, satisfied by a broad family of kicks, as confirmed by an exhaustive search over all admissible strings. The selection rule of the 2PK protocol is specific to the symmetry structure of the XXZ chain. It remains an open question whether a different criterion emerges when the Pauli strings themselves are sampled randomly \cite{Sun:2026wrk}. Since $c$ is, up to normalization, the lowest-order mixed free cumulant ($\kappa_2$) of $H$ and $PHP$ \cite{freeprobbook}, higher-order mixed cumulants \cite{Pappalardi:2022aaz} may provide sharper criteria for the selection of kicks and offer a physical interpretation through free probability theory \cite{voiculescu1986addition, Fava:2023pac}.

For the integrable boundary-defect chain, the protocol fails for every kick, with the first-order diagnostics remaining far from their design values. Design formation therefore needs both chaotic eigenvector statistics and the operator-space orthogonality between $H$ and $PHP$. Introducing a weak disorder into the longitudinal part of the Hamiltonian also enables a comparison (SM \ref{sm:Wstar}) with the recently proposed three-step protocol (3SP) \cite{Zhou:2026ubz}. In this regime, 2PK forms an approximate $k$-design, whereas the 3SP fails.

Our protocol uses a single Hamiltonian with a bulk defect and a fixed Pauli kick. So far, we have characterized the resulting doubly stochastic matrix $w$ through its moments up to fourth order. From the perspective of random doubly stochastic matrices \cite{Cappellini_2009}, its full eigenvalue distribution could reveal further structure and help identify which features are universal in the chaotic regime. It would also be interesting to study systems with constrained Hilbert spaces \cite{Bhore:2023tuc} and higher-spin generalizations of the XXZ chain \cite{Craps:2023rur}, where the underlying symmetries \cite{Marvian:2020sjz, Liu:2024vea} can further restrict or enlarge the set of admissible kicks and may lead to modifications of the selection rule in Eqs.\,\eqref{c_Delta_Ndw}-\eqref{Deltamax}. The 2PK ensemble is also closely related to temporal ensembles of Hilbert-space ergodicity \cite{Pilatowsky-Cameo:2024ljm, Mark:2024hrv}. Establishing this connection more explicitly would be useful. Finally, XXZ chains with defects can be realized in ultracold atoms in optical lattices \cite{Duan:2003xcq}, Rydberg-atom arrays \cite{Scholl:2021xeo}, and waveguide-QED platforms \cite{Tabares:2023bys}, making these systems promising candidates for experimental tests \cite{Vermersch:2018cke, Elben:2018xmb, Joshi:2020quh, Keenan:2022bzo, Joshi:2023rvd} of the protocol.


\section*{Acknowledgments}
We would like to thank Ivan M. Khaymovich for many fruitful discussions on related topics. S.S. is supported by the Swiss National Science Foundation under Division II (Grant No. 200020-219400). P.N. acknowledges the support from the ``COST Action CA22113: Short Term Scientific Mission (STSM), Theory Challenges 2026,'' hosted by the University of Geneva, where this work was initiated. P.N. is also supported by FWO-Vlaanderen projects G012222N and G0A2226N, and by the VUB Research Council through the Strategic Research Program in High-Energy Physics. Claude Opus 5 was used to assist with debugging and optimizing code originally written by the authors (in Mathematica and Python) and for grammatical corrections in the text. The authors take full responsibility for the scientific content and conclusions of this work. 

\bibliography{references}    

\clearpage
\onecolumngrid            

\begin{center}
  \textbf{\large Supplemental Material:\\[2pt]
  Designs without disorder: unitary $k$-designs from a single-site bulk defect}\\[8pt]
  Samudra Sur\,\,\href{https://orcid.org/0000-0002-4573-4038}
{\includegraphics[scale=0.05]{orcidid.pdf}}$^{1}$ and 
  Pratik Nandy\,\,\href{https://orcid.org/0000-0001-5383-2458}
{\includegraphics[scale=0.05]{orcidid.pdf}}$^{2,3}$\\[5pt]
  {\itshape\small $^{1}$Department of Quantum Matter Physics, University of Geneva, 24 quai Ernest-Ansermet, 1211 Genève, Switzerland}\\[3pt]
  {\itshape\small $^{2}$Theoretische Natuurkunde, Vrije Universiteit Brussel (VUB) and\\
  The International Solvay Institutes, Pleinlaan 2, B-1050 Brussels, Belgium}\\[3pt]
  {\itshape\small $^{3}$RIKEN Centre for Interdisciplinary Theoretical and Mathematical Sciences (iTHEMS), Wako, Saitama 351-0198, Japan}
\end{center}
\vspace{1em}

\setcounter{secnumdepth}{3}
\setcounter{section}{0}
\setcounter{equation}{0}
\setcounter{figure}{0}
\setcounter{table}{0}
\renewcommand{\thesection}{S\arabic{section}}
\renewcommand{\theequation}{S\arabic{equation}}
\renewcommand{\thefigure}{S\arabic{figure}}
\renewcommand{\thetable}{S\arabic{table}}

The Supplemental Material (SM) is organized as follows. Section \ref{sm:level_statistics} discusses the level-spacing statistics for boundary and bulk defect models. Section \ref{sm:derv} gives the derivation of the relation between the overlap ratio and the domain-wall number. Section \ref{sm:otherzz} shows the behavior of the frame potential for different $ZZ$ couplings; for six representative kicks in the boundary and bulk defect models (subsection \ref{sm:gapped}) and behavior beyond $J_{zz} > \sqrt{2}$ (subsection \ref{sm:beyond}). Section \ref{sm:Wstar} considers a weakly disordered Hamiltonian and compares the effectiveness of the 2PK protocol with a three-step protocol (3SP) of independently drawn Hamiltonians.

\section{Level spacing ratio of defect XXZ chains}
\label{sm:level_statistics}

\begin{figure}[h]
\hspace{-0.8cm}
\includegraphics[width=0.3\linewidth]{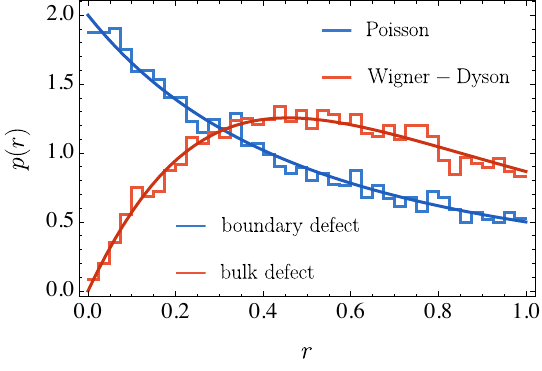}
\hspace{2ex}
\includegraphics[width=0.3\linewidth]{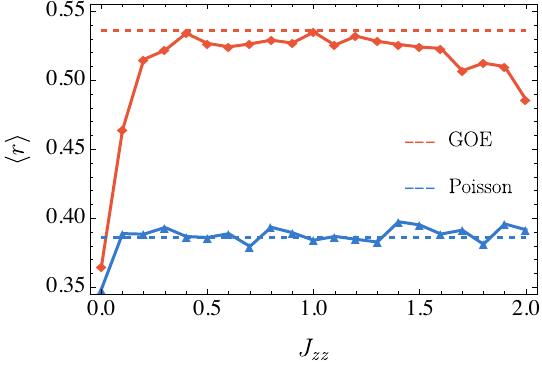}
    \caption{\textbf{Left:} Histogram of the $r$-value statistics for the XXZ chain with bulk and boundary defects. Statistics are computed within a single $S_{\mathrm{tot}}^z$ sector with all remaining symmetries broken. The system size is $L = 16$ with half-filling sectors of dimension $\binom{L}{L/2} = 12870$. Here $J_{zz} = 0.9$ with $(h_1, h_L) = (0.7,1.3)$ for the boundary and $h_{\lfloor L/2 \rfloor} = 1$ for the bulk perturbations. Solid lines are the analytic results for Poisson and Wigner-Dyson (WD) statistics. \textbf{Right:} $r$ value statistics with varying $J_{zz}$ for $L=14$.}
    \label{fig:rvalhistogram}
\end{figure}

To establish that the bulk defect drives the chain into the chaotic regime while boundary fields do not, we use the ratio of consecutive level spacings, which requires no unfolding. With $s_n := E_{n+1}-E_n$ the spacings of the ordered spectrum within a fixed magnetization sector, the $r$ ratio is defined \cite{Oganesyan:2007wpd, Atas2013distribution} as
\begin{align}
r_n = \frac{\min(s_n, s_{n-1})}{\max(s_n, s_{n-1})}\,, ~~~~ 0 \le r_n \le 1 \,,
\label{eq:rvalue}
\end{align}
whose mean takes the value $\langle r \rangle \simeq 0.386$ for uncorrelated
(Poisson) levels and $\langle r \rangle \simeq 0.5307$ for the GOE. The left panel of Fig.\,\ref{fig:rvalhistogram} shows the distribution $p(r)$ at $L=16$, where the half-filled sector has dimension $\binom{16}{8} = 12870$. The bulk-defect model follows the Wigner--Dyson form, including the level repulsion $p(r) \to 0$ as $r \to 0$, whereas the boundary model follows the Poisson form, consistent with its integrability \cite{Alcaraz:1987uk}. The right panel shows $\langle r \rangle$ across $J_{zz}$ at $L=14$: the boundary model remains at the Poisson value throughout, while the bulk-defect model rises to the GOE value by $J_{zz} \simeq 0.4$ and departs from it again in the strongly gapped regime, \emph{e.g.}, $J_{zz} = 1.5$. All spectra are resolved within a single magnetization sector, since mixing sectors would superpose uncorrelated spectra and spuriously drive $\langle r \rangle$ towards the Poisson value.

\section{Derivation of Equation (8).} \label{sm:derv}

In this section, we derive the main expression Eq.\,\eqref{c_Delta_Ndw}. The perturbed XXZ Hamiltonian is
\begin{align}
H=\sum_{i=1}^{L-1}
\left(X_iX_{i+1}+Y_iY_{i+1}
+J_{zz} Z_iZ_{i+1}\right)
+\sum_{j=1}^{L}h_jZ_j \equiv H_{XX} + H_{YY} +  H_{ZZ} + H_h\,,
\end{align}
where $H_{XX} := \sum_{i=1}^{L-1} X_i X_{i+1}$, $H_{YY} := \sum_{i=1}^{L-1} Y_i Y_{i+1}$, $H_{ZZ} := J_{zz} \sum_{i=1}^{L-1} Z_i Z_{i+1}$, and $H_{h} := \sum_{j=1}^{L} h_j Z_{j}$. Consider a longitudinal Pauli string $P=\prod_{j\in S}Z_j,
P^2=\mathds{1}$. The overlap ratio is defined as \cite{Nandy:2026zci}
\begin{align}
c\equiv \frac{\mathrm{Tr}(HPHP)}
{\mathrm{Tr}(H^2)}\,.
\end{align}
Using the conjugation of the single-site Pauli operator $ZXZ=-X, ZYZ=-Y, ZZZ=Z$, for a site $j$, we write
\begin{align}
P X_j P = (-1)^{\varepsilon_j}X_j\,,
\qquad
P Y_j P = (-1)^{\varepsilon_j}Y_j\,,
\qquad
P Z_j P = Z_j\,,
\end{align}
where $\varepsilon_j= 1$ for $j\in S$ and $0$ for $j\notin S$. In the indicator function notation, $\varepsilon_j = \textbf{1}[j \in S]$. Therefore, for an $XX$ bond, using $P^2 = \mathds{I}$, we have
\begin{align}
P X_iX_{i+1}P = (PX_iP)(PX_{i+1}P) =
(-1)^{\varepsilon_i+\varepsilon_{i+1}}
X_iX_{i+1} = (-1)^{|S\cap\{i,i+1\}|} X_iX_{i+1}\,, \label{cap}
\end{align}
where $| \cdot |$ is the cardinality of the set. Similarly, $P Y_iY_{i+1}P = (-1)^{|S\cap\{i,i+1\}|} Y_iY_{i+1}$. The exponent is odd when exactly one of the two sites $i$ and $i+1$ belongs to $S$, \emph{i.e.}, when the bond $(i,i+1)$ is a domain wall of the binary string $S$. Since $P$ is itself composed only of $Z$ operators,
\begin{align}
P Z_iZ_{i+1}P =Z_iZ_{i+1},~~~
P Z_jP =Z_j\,. \label{zcap}
\end{align}
Hence, the $ZZ$ interactions and the longitudinal fields are unchanged
under conjugation.

Now we define the domain-wall number as
\begin{align}
N_{\mathrm{dw}} = \sum_{i=1}^{L-1}
\mathbf{1} \left[
\left|S\cap\{i,i+1\}\right|=1
\right] =  \sum_{i=1}^{L-1}
\left|\varepsilon_i-\varepsilon_{i+1}
\right|\,.
\end{align}
Among the $L-1$ bonds, there are $N_{\mathrm{dw}}$ domain-wall
bonds and $L-1-N_{\mathrm{dw}}$ non-domain-wall bonds.

To evaluate the numerator of $c$, we expand $H=\sum_a \alpha_a Q_a$ in the Pauli-string basis $Q_a$, which satisfy the trace-orthogonality relation: $\mathrm{Tr}(Q_aQ_b)
= 2^L\delta_{ab}$. Therefore, when evaluating $\mathrm{Tr}(HPHP)$, all cross terms between distinct Pauli strings vanish. Thus, only the diagonal contributions from each term in $H$ survive.

Evaluating the contributions from the $XX$ bond, we have (using Eq.\,\eqref{cap})
\begin{align}
\mathrm{Tr}\left[(X_iX_{i+1})P(X_iX_{i+1})P\right] = (-1)^{|S\cap\{i,i+1\}|} \,\mathrm{Tr}(\mathds{1})
=2^L (-1)^{|S\cap\{i,i+1\}|}\,.
\end{align}
For each bond, the quantity $(-1)^{|S\cap\{i,i+1\}|}$ is $+1$ if neither site is in $S$, \emph{i.e.}, $|S\cap\{i,i+1\}|=0$; $-1$ if exactly one site is in $S$, \emph{i.e.}, $|S\cap\{i,i+1\}|=1$; and $+1$ if both sites are in $S$, \emph{i.e.}, $|S\cap\{i,i+1\}|=2$. Among the $L-1$ bonds, $N_{\mathrm{dw}}$ bonds contribute $-1$, while $L-1-N_{\mathrm{dw}}$ bonds contribute $+1$. Therefore, summing over $L-1$ bonds,
\begin{align}
    \mathrm{Tr}(H_{XX}PH_{XX}P) = \sum_{i=1}^{L-1} \mathrm{Tr}\left[(X_iX_{i+1})P(X_iX_{i+1})P\right] &= 2^L \sum_{i=1}^{L-1}
(-1)^{|S\cap\{i,i+1\}|} \nonumber \\
&= 2^L \left[(L-1-N_{\mathrm{dw}}) -
N_{\mathrm{dw}} \right] =
2^L\left[(L-1)- 2N_{\mathrm{dw}}\right].
\end{align}
Exactly the same argument gives $\mathrm{Tr}(H_{YY}PH_{YY}P)
= 2^L\left[(L-1)- 2N_{\mathrm{dw}}\right]$. The combined $XX+YY$ contribution is
\begin{align}
\mathrm{Tr}(H_{XX}PH_{XX}P) + \mathrm{Tr}(H_{YY}PH_{YY}P) =
2^{L+1}\left[(L-1)-2N_{\mathrm{dw}}\right]=
2^L\left[2(L-1)- 4N_{\mathrm{dw}}\right]\,.
\end{align}
Since the $ZZ$ terms are invariant under conjugation, using \eqref{zcap}, we have
\begin{align}
    \mathrm{Tr}(H_{ZZ}PH_{ZZ}P) = \sum_{i=1}^{L-1} J_{zz}^2 \mathrm{Tr} \left[Z_iZ_{i+1}Z_iZ_{i+1} \right] = 2^L J_{zz}^2(L-1)\,.
\end{align}
The longitudinal fields contribute to
\begin{align}
\mathrm{Tr}(H_hPH_hP) =
\sum_{j=1}^{L} h_j^2\,
\mathrm{Tr}(Z_j^2) =
2^L\sum_{j=1}^{L}h_j^2\,.
\end{align}
Combining the four contributions, we have
\begin{align}
\mathrm{Tr}(HPHP) &= \mathrm{Tr}(H_{XX}PH_{XX}P) + \mathrm{Tr}(H_{YY}PH_{YY}P) + \mathrm{Tr}(H_{ZZ}PH_{ZZ}P) + \mathrm{Tr}(H_hPH_hP) \nonumber \\ 
&=2^L\Big[2(L-1)-4N_{\mathrm{dw}}
+J_{zz}^2(L-1)+\sum_{j=1}^{L}h_j^2
\Big]\,.
\label{eq:numerator}
\end{align}
In the denominator, we find
\begin{align}
\mathrm{Tr}(H^2) = 2^L
\Big[ 2(L-1) + J_{zz} ^2(L-1)
+\sum_{j=1}^{L}h_j^2 \Big]\,.
\label{eq:denominator}
\end{align}
Taking the ratio of Eqs.\,\eqref{eq:numerator} and
\eqref{eq:denominator}, we obtain
\begin{align}
c(J_{zz},N_{\mathrm{dw}})=
\frac{\mathrm{Tr}(HPHP)}
{\mathrm{Tr}(H^2)} =
\frac{ 2(L-1)-4N_{\mathrm{dw}}
+J_{zz}^2(L-1)
+\sum_jh_j^2}{2(L-1)+J_{zz}^2(L-1)
+\sum_jh_j^2} =
1- \frac{4N_{\mathrm{dw}}}
{(2+J_{zz}^2)(L-1)+\sum_jh_j^2}\,.
\end{align}
which is Eq.\,\eqref{c_Delta_Ndw}.

\section{Designs with variation of ZZ coupling} \label{sm:otherzz}

\subsection{Representative kicks for $J_{zz} = 0.9$ and Frame potential for $J_{zz} = 1.1$} \label{sm:gapped}

\begin{table}[t]
\hspace{-0.25cm}
\small
\setlength{\tabcolsep}{2pt}
\begin{tabular}{cccc|cc|cc}
\hline\hline
Kick & String & weight & $N_{\mathrm{dw}}$ & \multicolumn{2}{c|}{$c_{\mathrm{bdy}}$} & \multicolumn{2}{c}{$c_{\mathrm{bulk}}$} \\
& & & & Eq.\,\eqref{c_Delta_Ndw} & Numerical & Eq.\,\eqref{c_Delta_Ndw} & Numerical \\
\hline
$P_1$ & ZIIIZIZIZZ & 5 & 6 & 0.126  & 0.111 & 0.087  & 0.081 \\
$P_2$ & ZIZIZIZIII & 4  & 7 & -0.019  & -0.037 & -0.065  & -0.072 \\
$P_3$ & IZIZIZIZII & 4  & 8 & -0.165  & -0.186 & -0.217  & -0.226  \\
$P_4$ & ZIZIZIZIZI  & 5  & 9 & -0.311  & -0.334 & -0.369  & -0.379 \\
$P_5$ & IIIIZIIIII & 1  & 2 & +0.709  & +0.704 & +0.696  & +0.694 \\
$P_6$ & ZZZZZIIIII & 5 & 1 & +0.854  & +0.852 & +0.848  & +0.847 \\
\hline\hline
\end{tabular}
\caption{The closed-form expression in Eq.\,\eqref{c_Delta_Ndw} is compared with the overlap ratio $c$ obtained numerically in the half-filled sector for the six representative Pauli kicks used in Fig.\,\ref{fig:KICKLISTgap} (left and middle), with $L=10$, $J_{zz}=0.9$, $(h_1,h_L)=(0.7,1.3)$ for the boundary-defect, and $h_{\lfloor L/2 \rfloor}=1$ for the bulk-defect models.}
\label{tab:kicks}
\end{table}

\begin{figure}[t]
\hspace*{-0.5cm}%
\includegraphics[width=0.6\linewidth]{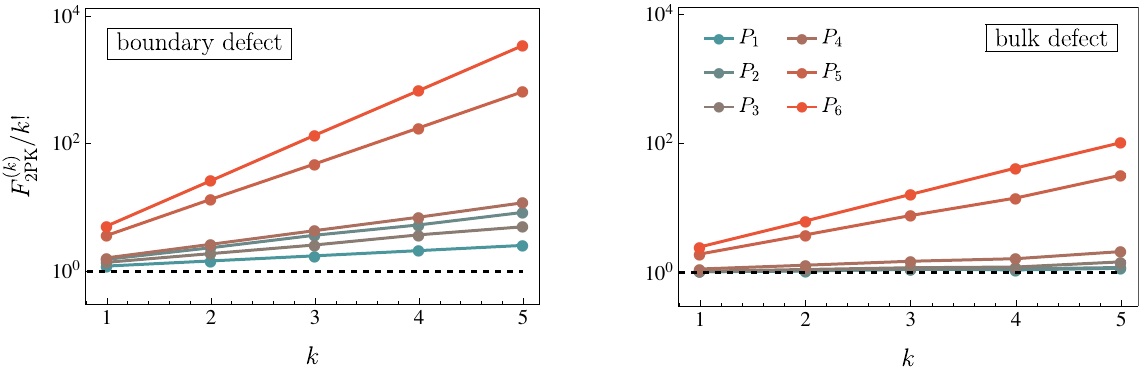}
\hfil
\includegraphics[width=0.285\linewidth]{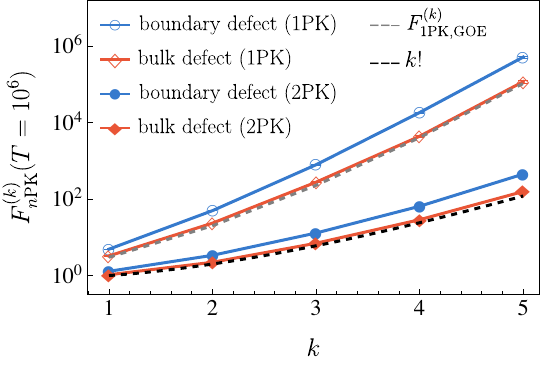}
\caption{\textbf{Left and Middle:} Normalized $F^{(k)}_{\mathrm{2PK}}/k!$ for the six Pauli kicks listed in Table \ref{tab:kicks} in the gapless regime $J_{zz} = 0.9$. \textbf{Right:} Frame potential $F^{(k)}$ for the XXZ chain with boundary and bulk perturbation under the 1PK and 2PK protocols in the gapped regime $J_{zz} = 1.1$. The chosen Pauli operator is ZIZIZIZIII. In all cases, the system size is $L = 10$ with a half-filling sector. The system parameters are  with $(h_1, h_L) = (0.7,1.3)$ for the boundary and $h_{\lfloor L/2 \rfloor} = 1$ for the bulk perturbation. The long-time limit is obtained by taking $T=10^6$, and the results are averaged over $10^5$ independent temporal realizations.}
\label{fig:KICKLISTgap}
\end{figure}

Table \ref{tab:kicks} compares the closed-form overlap ratio $c$ with values computed numerically in the half-filled sector for six representative kicks spanning the full range of domain-wall numbers. The agreement is at the $O(10^{-2})$ level, with the residual discrepancy arising from sector projection, under which distinct Pauli strings are no longer exactly trace-orthogonal. The optimal kicks are neither the highest- nor the lowest-weight strings. For example, $P_6$ has weight $5$ but only one domain wall and performs among the worst, while $P_5$ has weight $1$ and performs similarly poorly. Thus, the spatial structure of the kick relative to the terms in the Hamiltonian determines its effectiveness in design formation. The left and middle panels of Figure \ref{fig:KICKLISTgap} show the normalized frame potential for six representative kicks in the gapless regime, $J_{zz}=0.9$. For the bulk-defect model, most of the kicks lead to approximate $k$-designs, whereas for the boundary-defect model none of them does.

The right panel of Fig.\,\ref{fig:KICKLISTgap} demonstrates the behavior of frame potential in the gapped regime $J_{zz} = 1.1$, within the bound \eqref{Deltamax}. The behavior is similar to the gapless regime in Fig.\,\ref{fig:FPnPKXXZL10Jzz0p9Tplot}.

\subsection{Results beyond $J_{zz} > (J_{zz})_{\mathrm{max}}$} \label{sm:beyond}

\begin{figure}[t]
\hspace{-0.8cm}
\includegraphics[width=0.7\linewidth]{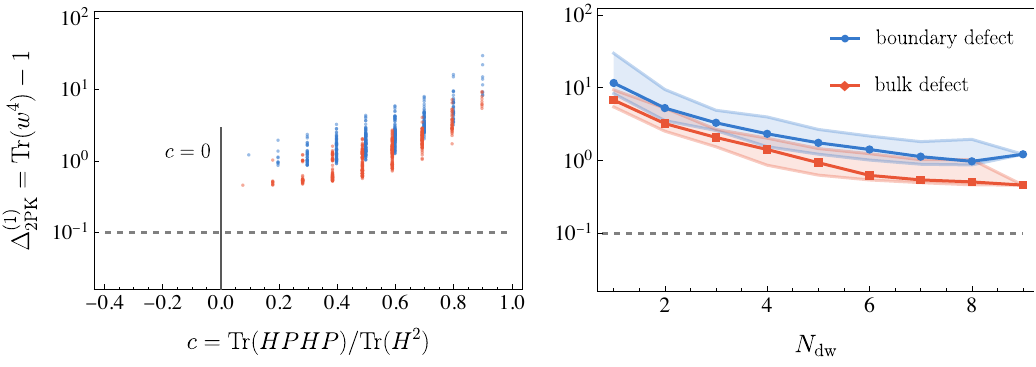}
\hfill
\includegraphics[width=0.36\linewidth]{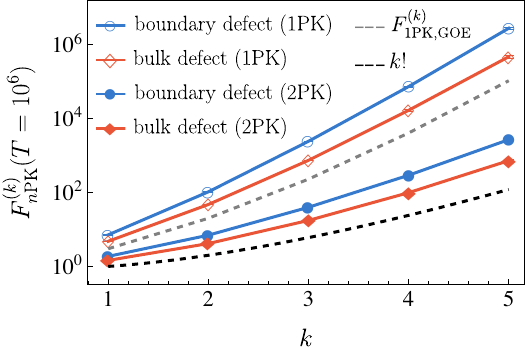}
    \caption{\textbf{Top, Left:} Selection rule for the Pauli kicks for $L=10$, $J_{zz} = 1.5$, $(h_{1}, h_L) = (0.7,1.3)$ (boundary defect) and $h_{\lfloor L/2 \rfloor} = 1$ (bulk defect). \textbf{Top, Right:} Variation of $\Delta_{\mathrm{2PK}}^{(1)}$ with $N_{\mathrm{dw}}$. Here, no optimal kick is found. \textbf{Bottom:} The 2PK frame potential with the choice of the best possible kicks. The bulk-defect model does not lead to the approximate $k$-design. Here $T = 10^6$ and a total of $10^5$ temporal samples are taken.}
    \label{fig:AppL10Jzz1p5}
\end{figure}

The main text uses $J_{zz}=0.9$ at $L=10$ for the exhaustive kick search. This choice is motivated by the fact that, in the bulk-defect model, a large number of kicks satisfy $\Delta^{(1)}_{\mathrm{2PK}}\leq 0.1$, as shown in Fig.\,\ref{fig:FP2PKGUEwithbetaplot}. Although Eq.\,\eqref{Deltamax} shows that $c=0$ is not exactly reachable for $J_{zz}>(J_{zz})_{\mathrm{max}}$, the best admissible kicks still have small $|c|$. The failure is therefore due to the departure from GOE eigenvector statistics in the gapped regime. To see this explicitly, Fig.\,\ref{fig:AppL10Jzz1p5} (top-left and top-right panels) shows the exhaustive kick search at $J_{zz}=1.5$. In this case, for both the boundary- and bulk-defect models, we find $\min \Delta^{(1)}_{\mathrm{2PK}}\sim O(1)$, indicating that an approximate design does not form.  

In the bottom panel of Fig.\,\ref{fig:AppL10Jzz1p5}, we numerically compute the frame potential for $k \leq 5$, with $T = 10^6$ for the kick $P$ = IZIZIZIZZI (for boundary-defect) and $P$ = ZIZIZIZIZI (for bulk-defect). This kick is chosen by the minimum of $\Delta^{(1)}_{\mathrm{2PK}}$ for all the admissible kicks.

\section{Disorder threshold for the 2PK selection rule and Comparison with 3SP}
\label{sm:Wstar}

\begin{figure}[t]
\hspace{-0.8cm}
\includegraphics[width=0.7\linewidth]{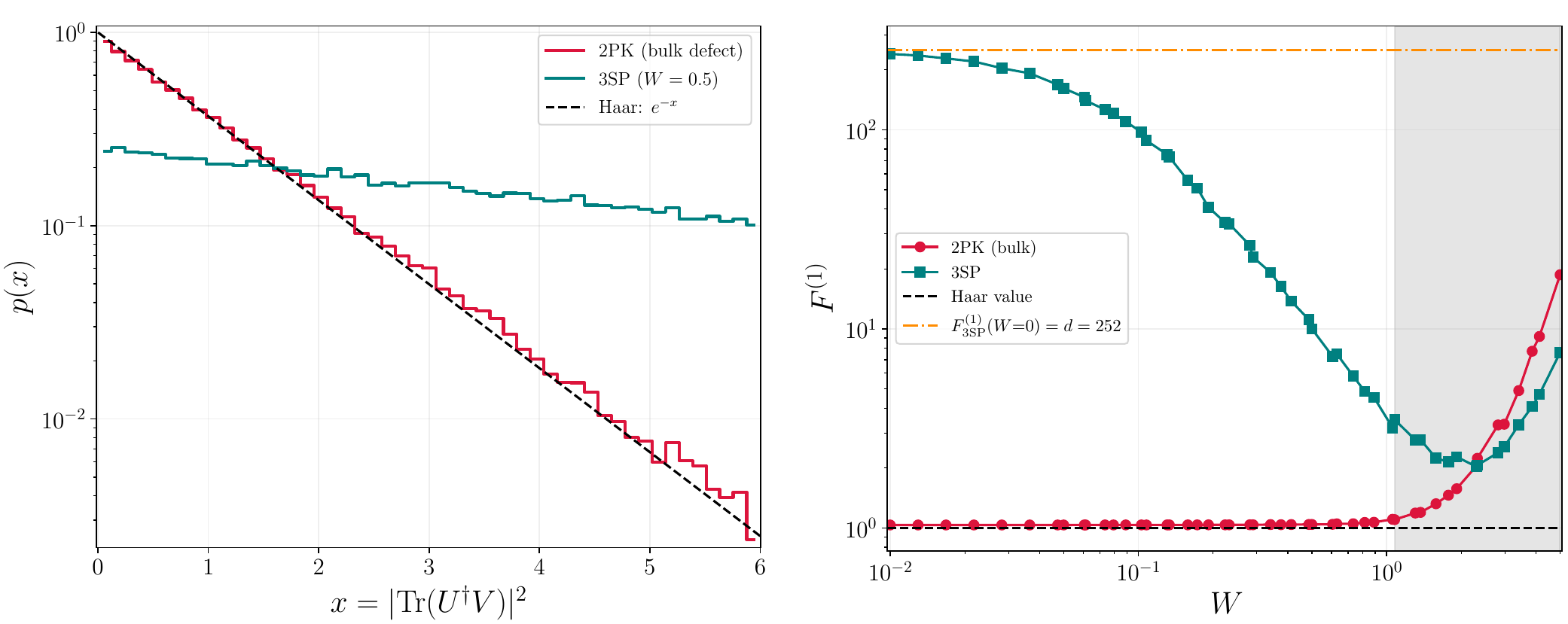}
    \caption{\textbf{Left:} Distribution of the pairwise overlap $x=|\mathrm{Tr}(U^\dagger V)|^2$ at $L=10$, against the Haar law $e^{-x}$ (dashed), whose moments $\langle x^k\rangle=k!$ reproduce the Haar frame potentials at all $k$ simultaneously. Here $10^5$ temporal samples are taken. \textbf{Right:} First-order frame potential versus disorder strength $W$ for both protocols. We have used the exact $T \to \infty$ closed form for the Frame potential. However, each value is averaged over 50 independent disorder realizations; within each realization, the Hamiltonian(s) are drawn once and held fixed, with randomness entering only through the evolution times. The shaded region is where \eqref{eq:fieldbound} does not admit any optimal kick. Dashed line: Haar value $F^{(1)}=1$. The kick is $P_1$ from Table \ref{tab:kicks}.}
    \label{fig:momentXXZ}
\end{figure}

We compare the 2PK protocol with the three-step protocol (3SP) \cite{Zhou:2026ubz}, where the ensemble is built from three successive quenches $\mathcal{E}_{\mathrm{3SP}}=\big\{U_{\mathrm{3SP}}=e^{-iH_3t_3}\,e^{-iH_2t_2}\,e^{-iH_1t_1}\big\}$ with $H_1,H_2,H_3$. Here the Hamiltonians $H_1,H_2,H_3$ are drawn independently from a disordered ensemble and then fixed. The randomness enters only through the temporal sampling of the evolution times.
As a consequence, we choose the Hamiltonians to have random disorder along with the defect terms. More precisely, $H_i$ have the structure 
\begin{align}
    H = \sum_{i} (X_iX_{i+1} + Y_iY_{i+1}+ J_{zz}\, Z_i Z_{i+1}) + \sum_{i}h_i Z_i +\sum_{i} \delta_i Z_i\,,
\end{align}
where $h_i$ is taken to be the bulk defect acting only at $i= \lfloor L/2 \rfloor$, and $\delta_i \in [-W, W]$. The disorder strength $W$ is set at $0.5$, and the number of temporal samples used is $10^5$. We also choose $J_{zz} = 0.9$. For the 2PK protocol, the Pauli kick is chosen to be $P_1$ from the table \ref{tab:kicks}.

Figure \ref{fig:momentXXZ} (left panel) shows the resulting overlap distributions. 2PK on the bulk-defect chain follows the distribution $e^{-x}$, consistent with the Haar prediction. On the other hand, 3SP carries far more weight at large $x$, \emph{i.e.}, its members remain mutually correlated. Figure \ref{fig:momentXXZ} (right panel) shows exact $T \to \infty$ limit of $F^{(1)}$ across $W$. In the limit of $W \to 0$, 3SP has a particularly simple form. The three Hamiltonians in 3SP become identical, so the evolution is given by a single global unitary operator with randomized times
\begin{align}
    U_{\mathrm{3SP}} = e^{- i H_3 t_3} e^{- i H_2 t_2} e^{- i H_1 t_1} \xrightarrow[W \to 0]{} e^{- i H t}\,, ~~~ t := t_1 + t_2 + t_3\,.
\end{align}
Consequently, the frame potential is simply (with $t' := t_1' + t_2' + t_3'$)
\begin{align}
    F^{(k)}_{\mathrm{3SP}, W = 0} = \mathbb{E}_{t, t'} \Big[|\mathrm{Tr} (U(t')^{\dagger} U(t))|^{2k} \Big] = \mathbb{E}_{t, t'}\Big[|\mathrm{Tr} e^{-i H (t - t')}|^{2k} \Big] = \mathbb{E}_{t, t'}\Big[K(t - t')^k  \Big]\,,
\end{align}
where $K(\tau) =|\mathrm{Tr} (e^{-i H \tau})|^2 = \sum_{m,n} e^{-i (E_m - E_n) \tau}$ is the unnormalized spectral form factor (SFF). Focusing on $k=1$ and expanding the sum, we get
\begin{align}
F^{(1)}_{\mathrm{3SP},W=0} = \mathbb{E}_{t,t'} \left[ \sum_{m,n}
e^{-i(E_m-E_n)(t-t')} \right] = \sum_{m,n} \mathbb{E}_{t,t'}
\left[ e^{-i(E_m-E_n)(t-t')}
\right] = \sum_{m,n} |\phi(E_m - E_n)|^2\,,
\end{align}
where $\phi(\omega) = \mathbb{E}_{t}[e^{-i\omega t}]$ is the temporal characteristic function of the total evolution time $t = t_1 + t_2 + t_3$. Each energy gap $E_m-E_n$ is weighted by the squared modulus of the characteristic function $\phi(E_m - E_n)$. Since the three times are drawn independently and uniformly, \emph{i.e.}, $t_i \sim \mathrm{Uniform}[0,T]$, the characteristic function factorizes
\begin{align}
\phi(\omega) = [\phi_1(\omega)]^3\,,~~~
\phi_1(\omega) = \frac{1}{T}\int_0^T e^{-i\omega t}\,dt
= e^{-i\omega T/2}\,
\frac{\sin(\omega T/2)}{\omega T/2}\,.
\end{align}
The overall phase factor disappears in the squared modulus, leading to
\begin{align}
F^{(1)}_{\mathrm{3SP},W=0} = \sum_{m,n}
\operatorname{sinc}^6
\left(\frac{(E_m-E_n)T}{2}\right) = \sum_{m = 1}^d \mathbf{1} + \sum_{m \neq n} \operatorname{sinc}^6
\left(\frac{(E_m-E_n)T}{2}\right)\,,
\end{align}
where $\operatorname{sinc}(x):=\sin(x)/x$ is the sinc function. For a nondegenerate spectrum, the first term is the diagonal terms $m=n$, which contribute to a factor $d$. The second term is the off-diagonal terms $m \neq n$, which vanish as $T \to \infty$. Hence, we obtain
\begin{align}
F^{(1)}_{\mathrm{3SP},W=0}
\xrightarrow[T\to\infty]{}
d\,. \label{ag}
\end{align}
This can also be generalized to higher $k$ as
\begin{align}
    F^{(k)}_{\mathrm{3SP},W=0} =  \sum_{\mathbf{m,n}} |\phi(\Omega_{\mathbf{m,n}})|^2\,,~~~~ \Omega_{\mathbf{m,n}} = \sum_{a=1}^k E_{m_a} - \sum_{a=1}^k E_{n_a}\,,
\end{align}
where $\Omega_{\mathbf{m,n}}$ is the difference between the two sums of $k$ energy levels or resonance detuning. The exact $k$-th order resonance implies $\Omega_{\mathbf{m,n}} = 0$.

In the $T \to \infty$ limit, an analytic expression for the frame potential can be obtained. It is given by
\begin{align}
F^{(k)}_{\mathrm{3SP},W=0} \xrightarrow[T\to\infty]{} \sum_{\substack{r_1+\cdots+r_d=k \\ r_i\geq 0}}\left(\frac{k!}{r_1!\cdots r_d!}\right)^2 = (k!)^2 [x^k] I_0(2 \sqrt{x})^d\,,
\end{align}
where $[x^k] f(x)$ denotes the coefficient of $x^k$ in $f(x)$ and $I_0$ is the zero-th order modified Bessel function of the first kind. For $k=1, 2, 3$ we can check
\begin{align}
F^{(1)}_{\mathrm{3SP},W=0}
\xrightarrow[T\to\infty]{}
d\,,~~~~ F^{(2)}_{\mathrm{3SP},W=0}
\xrightarrow[T\to\infty]{}
2d^2 - d\,,~~~~ F^{(3)}_{\mathrm{3SP},W=0}
\xrightarrow[T\to\infty]{} 6 d^3 - 9 d^2  + 4d\,,
\end{align}
agreeing with the $k = 1$ result in \eqref{ag}. For fixed $k$ and large $d$, we have
\begin{align}
    F^{(k)}_{\mathrm{3SP},W=0} = k! d^k - \frac{k! k(k-1)}{4} d^{k-1} + O(d^{k-2})\,.
\end{align}
Therefore, the leading term is $k! d^k$ compared to the Haar value $k!$. All the above results hold for 1SP (one-step protocol) or 0PK (zero Pauli-kick) protocol. 

In Fig.\,\ref{fig:momentXXZ} (right panel), the 3SP frame potential decreases from $F^{(1)}\to d$ at $W\to0$ to a shallow minimum near $W\simeq2$. It increases again as the eigenstates localize, but it never reaches the Haar value over the range of $W$ studied. In contrast, the 2PK ensemble remains at $F^{(1)} \simeq 1$ up to $W \simeq 1$, indicating that disorder has little effect on the ensemble generated by a single fixed Hamiltonian. It degrades at larger $W$, where the growing longitudinal fields drive the chain toward localization and the eigenvector statistics cease to be Gaussian. Independently, for $W > W_{*} \simeq 1.71$, no admissible kick satisfies $c = 0$. Eq.\,\eqref{Deltamax} ceases to admit an optimal kick once $\sum_jh_j^2>(2-J_{zz}^2)(L-1)$, which defines $W_{*}$ (see the derivation in the next paragraph).

Setting $c=0$ in Eq.\,\eqref{c_Delta_Ndw} fixes the optimal domain-wall count, \emph{i.e.}, Eq.\,\eqref{Ndwstar},
\begin{align}
N^{*}_{\mathrm{dw}}=\frac{1}{4}\Big[(2+J_{zz}^2)(L-1)+\sum_j h_j^2\Big]\,,
\label{eq:Ndwstar-SM}
\end{align}
while the open chain contains $L-1$ bonds, so $N_{\mathrm{dw}}\le L-1$, the maximum being attained by the two N\'eel strings. An optimal kick therefore exists if and only if $N^{*}_{\mathrm{dw}}\le L-1$, \emph{i.e.},
\begin{align}
\sum_j h_j^2\;\le\;(2-J_{zz}^2)(L-1)\,.
\label{eq:fieldbound}
\end{align}
The above equation is equivalent to Eq.\,\eqref{Deltamax} as a constraint on the fields at fixed anisotropy. This follows because the Ising term and the longitudinal fields commute with every admissible ($U(1)$-preserving) kick and hence remain unchanged. Only the $XX$ and $YY$ bonds can change sign, and this must be at least half of $\mathrm{Tr}(H^2)$ for equipartition to be possible.

For the disordered bulk-defect chain, $h_{\lfloor L/2 \rfloor} = 1+\delta_{\lfloor L/2\rfloor}$
and $h_j=\delta_j$ otherwise, with $\delta_j$ uniform on $[-W,W]$ so that
$\overline{\delta_j^2}= \frac{1}{2W}\int_{-W}^W \delta^2 d \delta =  W^2/3$. Hence
\begin{align}
\sum_j h_j^2 \;\simeq\; 1+\frac{L}{3}W^2\,,
\end{align}
and \eqref{eq:fieldbound} is saturated at
\begin{align}
W_{*}=\sqrt{\frac{3\big[(2-J_{zz}^2)(L-1)-1\big]}{L}}\,,
\label{eq:Wstar}
\end{align}
which gives $W_{*} \simeq 1.71$ for $L=10$ and $J_{zz}=0.9$. For $W>W_{*}$, no admissible kick satisfies the selection rule. This is consistent with the result of Fig.\,\ref{fig:momentXXZ}, where the 2PK frame potential is flat through $W=1.5$ and departs between $W=1.5$ and $W=3$. For $W \leq 1$, 3SP built from three weakly disordered realizations (disorder is added in the longitudinal field) does not reach the Haar value, whereas 2PK with a single realization does.

\end{document}